# High-performance silicon-graphene hybrid plasmonic waveguide photodetectors beyond 1.55 μm

Jingshu Guo[1,5], Jiang Li[1,5], Chaoyue Liu[1], Yanlong Yin[1], Wenhui Wang[3], Zhenhua Ni[3], Zhilei Fu[4], Hui Yu[4], Yang Xu[2,4], Yaocheng Shi[1,2], Yungui Ma[1], Shiming Gao[1,2], Liming Tong[1], Daoxin Dai[1,2]*

1. State Key Laboratory for Modern Optical Instrumentation, College of Optical Science and Engineering, Zhejiang University, Zijingang Campus, Hangzhou, China.
2. Ningbo Research Institute, Zhejiang University, Ningbo 315100, China.
3. Department of Physics and Key Laboratory of MEMS of the Ministry of Education, Southeast University, Nanjing 211189, China.
4. College of Information Science and Electronic Engineering, Zhejiang University, Hangzhou, Zhejiang, 310027, China.
5. These authors contributed equally to this work: Jingshu Guo, Jiang Li.

*Corresponding author: dxdai@zju.edu.cn

**Abstract:** A fast silicon-graphene hybrid plasmonic waveguide photodetectors beyond 1.55 μm is proposed and realized by introducing an ultra-thin wide silicon-on-insulator ridge core region with a narrow metal cap. With this novel design, the light absorption in graphene is enhanced while the metal absorption loss is reduced simultaneously, which helps greatly improve the responsivity as well as shorten the absorption region for achieving fast responses. Furthermore, metal-graphene-metal sandwiched electrodes are introduced to reduce the metal-graphene contact resistance, which is also helpful for improving the response speed. When the photodetector operates at 2 μm, the measured 3dB-bandwidth is >20 GHz (which is limited by the experimental setup) while the 3dB-bandwith calculated from the equivalent circuit with the parameters extracted from the measured $S_{11}$ is as high as ~100 GHz. To the best of our knowledge, it is the first time to report the waveguide photodetector at 2 μm with a 3dB-bandwidth over 20 GHz. Besides, the present photodetectors also work very well at 1.55 μm. The measured responsivity is about 0.4 A/W under a bias voltage of −0.3 V for an optical power of 0.16 mW, while the measured 3dB-bandwidth is over 40 GHz (limited by the test setup) and the 3 dB-bandwidth estimated from the equivalent circuit is also as high as ~100 GHz, which is one of the best results reported for silicon‐graphene photodetectors at 1.55 μm.

## 1. Introduction

Silicon photonics has been recognized as a very promising technology for many applications because of the unique advantages[1], e.g., the CMOS compatibility, the high integration density, etc. Currently, silicon photonics has been developed very successfully for the applications of optical interconnects operating with the near-infrared wavelength-band of 1.31/1.55 μm. In order to satisfy the increasing demands in optical communication[2], nonlinear photonics[3], lidar[4], and optical bio-sensing[5,6], it is desired to develop silicon photonic devices working beyond the wavelength-band of 1.55 μm (e.g., 2 μm). Among various photonic devices, high-speed waveguide photodetectors are one of the most important elements and always play very important roles in various photonic systems. As silicon is transparent in the wavelength range from 1.1 μm to 6 μm, another material with high absorption of light is usually necessary to be introduced for photodetection. Previously, high-performance waveguide photodetectors on silicon for 1.3-1.55 μm have been realized successfully by introducing some mature semiconductor material like germanium[7], however, which does not work well for the wavelength-band beyond 1.55 μm.

In order to realize photodetectors beyond 1.55 μm, GeSn was introduced by using the epitaxy growth process[8,9] and recently a surface-illuminated GeSn/Ge quantum-well photodiode on silicon was reported with a bandwidth of 10 GHz as well as a responsivity of 0.015 A/W when operating at 2 μm[10]. However, the wavelength-band is still limited (typically <2.3 μm) and there is no results for >10 GHz Si/GeSn waveguide photodetectors yet. As an alternative, GaInAsSb/GaAsSb semiconductor materials with quantum-wells were introduced[11] and hybrid silicon/III-V waveguide photodetectors for the wavelength-band around 2.3 μm were realized by using the adhesive BCB bonding process. In this case, the fabrication is still quite complicated. More recently, a monolithic silicon photodiode with a 3dB-bandwidth of 15 GHz for the 2 μm wavelength-band was demonstrated by utilizing the defect-level absorption in silicon[12]. One might notice that the bias voltage is as high as ~30 V and further extension to longer wavelengths is hard.

In contrast, two-dimensional materials provide a new and promising option for enabling active photonic devices on silicon[13-16]. In particular, currently graphene[14,15] and black-phosphorus (BP)[16-19] have been popular for realizing waveguide photodetectors on silicon. Recently, we reported a silicon-BP hybrid waveguide photodetector with a 3 dB-bandwidth of 1.33 GHz and a high-speed operation of 4.0 Gbps as well as high responsivity of ~0.3 A/W at 2 μm[19]. It is noted that the bandwidth of the reported BP photodetectors is still limited, e.g., less than 3 GHz[17-19]. In addition, the fabrication is still not easy because the large-size high-quality BP sheet is not available yet and the BP sheet needs some special protection. In contrast, large-size graphene sheets are commercially available and can be transferred/patterned easily when working together with silicon photonics. As one of the most popular 2D-materials for silicon photonics, graphene has some unique advantages as follows[13,14]. First, it has very high mobility enabling fast responses of photodetection. Second, it has an ultra-broad light absorption wavelength-band covering the near-infrared range as well as the wavelength-band beyond 1.55 μm. Third, there is no material and structure mismatch when integrated with silicon photonic circuits. With these advantages, in recent years many exciting results for silicon-graphene photodetectors were reported for the 1.31/1.55 μm wavelength-band. For example, silicon-graphene waveguide photodetectors operating at 1.55 μm were realized with a bandwidth of >76 GHz successfully in a 6-inch wafer process line[20]. More recently silicon-graphene waveguide photodetectors working at 1.55 μm were reported with a bandwidth even as high as 100 GHz[21-23]. However, the responsivity performance of these high-speed graphene waveguide photodetectors still needs improvements when the low bias voltages are applied in case of large dark currents. For example, the graphene waveguide photodetectors utilizing the photovoltaic (PV) effect[23-29], the photo-thermoelectric (PTE) effect[30-34], or the internal photoemission effect (IPE)[35] usually have limited responsivities (<78 mA/W at zero bias[30], and <170 mA/W @ <0.4 V[32]). Alternatively, another monolayer-graphene photodetector utilizing the bolometric (BOL) effect was reported with a high responsivity of 500 mA/W when operating with a bias voltage of −0.4 V and an input power of 0.08 mW. However, the responsivity is reduced to 200 mA/W for a high input power e.g. >0.6 mW[21]. In addition, a $MoS_2$/graphene/BN/graphene tunneling photodiode was recently reported with boht low dark current (on nA scale), high responsivity of 0.24 A/W and estimated bandwidth of 56 GHz, but the bias voltage is too large 10V[36]. For the wavelength-band beyond 1.55 μm, some surface-illuminated grapheme photodetectors were demonstrated[37-42]. However, in this case the responsivity is low because the light absorption is quite limited, as it is well known. In recent years, a few silicon-graphene waveguide photodetectors were realized[43-45] and unfortunately the measured bandwidths are several hundreds of KHz or less. To the best of our knowledge, currently there are not high-speed (e.g., >10 GHz) silicon-graphene waveguide photodetectors reported for the mid-infrared range beyond the wavelength-band of 1.55 μm.

In this paper, we propose and demonstrate ultrafast silicon-graphene hybrid plasmonic waveguide photodetectors beyond 1.55 μm. The present hybrid plasmonic waveguide consists of an ultra-thin wide silicon-on-insulator (SOI) ridge core region with a narrow metal cap, between which there is an ultra-thin $Al_2O_3$ insulator layer. With such a hybrid plasmonic waveguide, the light absorption in graphene is enhanced while the metal absorption loss is reduced simultaneously, which helps greatly improve the responsivity. When operating at 2 μm, the present silicon-graphene hybrid plasmonic waveguide photodetector has a responsivity of ~70 mA/W for an optical power of 0.28 mW. Here, the metal-graphene-metal sandwiched electrodes[46] are introduced to reduce the metal-graphene contact resistance, and thus improve the response speed of the photodetectors. For the present silicon-graphene photodetectors, the measured 3 dB-bandwidth is higher than 20 GHz (which is limited by the experimental setup) while the 3dB-bandwith calculated from the equivalent circuit with the parameters extracted from the measured $S_{11}$ is as high as ~123 GHz. To the best of our knowledge, it is the first time to report the waveguide photodetector at 2 μm with a measured 3dB-bandwidth over 20 GHz. Meanwhile, the present silicon-graphene hybrid plasmonic waveguide photodetectors also work well when operating at 1.55 μm. The responsivity is about 0.4 A/W under a bias voltage of −0.3 V for an optical power of 0.16 mW, while the measured 3 dB-bandwidth is over 40 GHz (still limited by the test setup) and the 3 dB-bandwith estimated from the equivalent circuit is as high as ~94 GHz, which is one of the best results reported for silicon-graphene hybrid photodetectors at 1.55 μm.

2. **Structure and Design.**

Figures 1(a)-(b) show the configuration of the present silicon-graphene hybrid plasmonic waveguide photodetector, which consists of a passive input section based on an SOI strip waveguide and an active region based on a silicon-graphene hybrid plasmonic waveguide. These two parts are connected through a mode converter based on a lateral taper structure. As shown in Fig. 1(c), the present hybrid plasmonic waveguide has an SOI ridge core region, an ultra-thin $Al_2O_3$ insulator layer, a graphene sheet, and a narrow metal cap. The SOI ridge height is chosen as small as 50 nm,

which is helpful to avoid the damage of the graphene sheet during the fabrication processes. The narrow metal cap at the middle is used as the signal electrode, while the ground electrodes are placed far away from the SOI ridge to avoid high metal absorption loss. In particularly, here we introduce the metal-graphene-metal sandwiched structure for the ground electrodes to achieve reduced graphene-metal contact resistances, which helps achieve a large 3 dB-bandwidth. In order to enhance the light absorption of graphene in the active region, the silicon core-layer is chosen as thin as 100 nm (instead of the regular thickness of 220 nm[20,30-31]) for the silicon-graphene hybrid plasmonic waveguide. In addition, the metal strip (the signal electrode) on the top of the silicon ridge also helps achieve the field localization for further enhancing the graphene absorption. One might notice that hybrid plasmonic waveguides usually have high propagation losses due to the metal absorption, which does not have any contribution to the generation of the photocurrent and thus prevents to achieve high responsivity. For conventional silicon-graphene hybrid plasmonic waveguide photodetectors, the width of the metal strip is not much smaller than the silicon core width[26]. As a result, the undesired metal absorption is even higher than the desired graphene absorption and thus the responsivity is usually low.

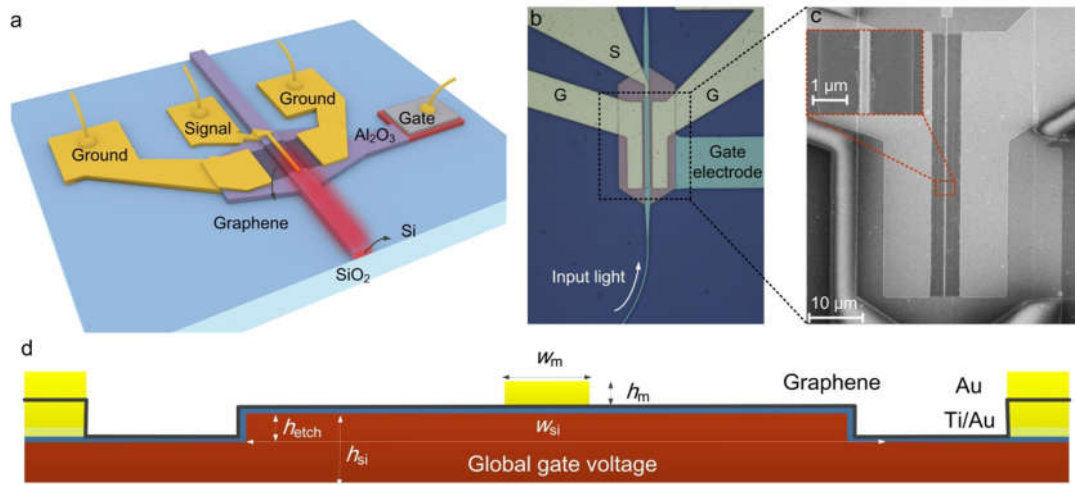

Fig. 1. The present silicon-graphene hybrid plasmonic waveguide photodetector. (a) Schematic configuration. (b) Optical microscope and (c) SEM pictures. (d) Cross-section of the present silicon-graphene hybrid plasmonic waveguide with the signal electrode at the middle and the ground electrodes at both sides (here the metal-graphene-metal sandwich structure is utilized).

In this paper, our silicon-graphene hybrid plasmonic waveguide is designed with a wide silicon ridge on a thin silicon platform, as shown in Fig.1(c). In this way, even when the metal strip at the middle is designed to be wide sufficiently for achieving a low metal-graphene contact resistance, the metal absorption can be minimized and the light absorption of graphene is still dominant compared to the metal absorption. In order to be able to manipulate the graphene chemical potential in the whole channel, a gate voltage is applied to the gate electrode on the top of the slab region of the SOI ridge waveguide, as shown in Fig. 1(a). Figures 2(a)-(b) show the calculation results for evaluating the light absorption induced by the graphene sheet and the metal strip in the present silicon-graphene hybrid plasmonic waveguide as the waveguide dimensions vary. Here the finite-element method mode-solver tool (COMSOL) was used (see more details in Supplementary Section 1). The graphene absorptance is given by $\eta(L)=\eta_g(1-10^{-0.1\alpha L})$, where $L$ is the propagation distance, $\alpha$ is the mode absorption coefficient in dB/μm, $\eta_g$ is the ratio of the graphene absorption to the total absorption, i.e., $\eta_g = \frac{\alpha_g}{\alpha} = \frac{\alpha_g}{\alpha_g+\alpha_m}$ (here $\alpha_g$ and $\alpha_m$ are respectively the absorption coefficients of the graphene sheet and the metal strip). Since only the graphene absorption contributes to the photocurrent, one should maximize the ratio $\eta_g$ so that the graphene absorption is more dominant than the metal absorption in order to improve the responsivity. Fig. 2(a) shows the absorption ratio $\eta_g$ and the results for the absorption coefficients ($\alpha_g$, $\alpha_m$) as the ridge width $w_{si}$ varies from 0.5 μm to 4.0 μm. Here the width and the height of the metal strip are chosen as $w_m$=200 nm and $h_m$=50 nm. As shown in Fig. 2(a), it can be seen that the graphene absorption ratio $\eta_g$ becomes higher when choosing a wider ridge. When the ridge width $w_{si}$ is chosen to be larger than 3μm, the ratio $\eta_g$ is higher than 70%. Meanwhile, it is noted that the absorption coefficients ($\alpha_g$, $\alpha_m$) becomes lower when choosing a wider ridge, which is simply due to more optical confinement in the silicon region and less light-matter interaction in the absorption regions. As a result, one needs to choose a longer absorption length for sufficient absorption in the photodetector, which prevents to achieve fast responses regarding the RC-constant limitation. Fortunately, the light absorption can be enhanced greatly by reducing the silicon core height $h_{Si}$, as shown in Fig. 2(a), where the absorption coefficients ($\alpha_g$, $\alpha_m$) for the cases with different silicon core heights $h_{Si}$=220, 160, and 100 nm are given. From this figure, it can be seen that the absorption coefficients $\alpha_g$ and $\alpha_m$ increase by more than 100% when the core height $h_{Si}$ is reduced from 220 nm to 100 nm. This is attributed to the stronger evanescent field for the case with a thinner silicon core. Meanwhile, the graphene absorption ratio $\eta_g$ increases slightly as the core height $h_{Si}$ decreases. As a result, an ultrathin silicon core is preferred to achieve strong light absorption so that one can achieve a short the absorption section. Here we choose $h_{Si}$ = 100 nm for our devices regarding the feasibility of the fabrication processes. In order to avoid long carrier transit time between the electrodes, the ridge width is chosen as $w_{Si}$ = 3 μm. With this design, the absorption coefficients are $(\alpha_g, \alpha_m)$ = (0.230, 0.098) dB/μm, and the graphene absorption ratio $\eta_g$ is about 70%.

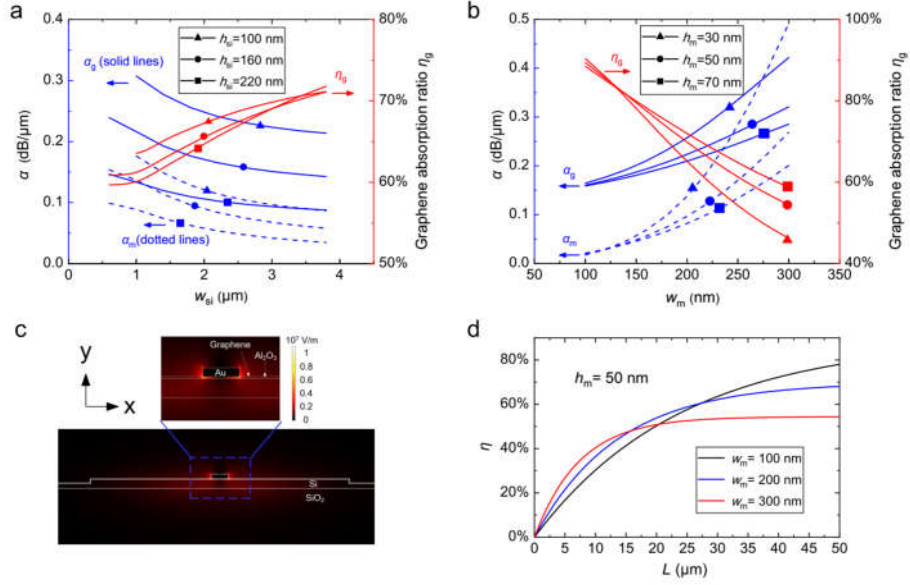

Fig. 2 (a) Calculated absorption coefficients $(\alpha_g, \alpha_m)$, and the graphene absorption ratio $\eta_g$ as the silicon ridge width $w_{si}$ varies for the cases with different silicon ridge heights $h_{si}$. Here $w_m$ = 200 nm, and $h_m$ = 50 nm. (b) Calculated absorption coefficients $(\alpha_g, \alpha_m)$, and the graphene absorption ratio $\eta_g$ as the metal strip width $w_m$ varies for the cases with different metal heights $h_m$. Here $w_{si}$ = 3 μm, and $h_{si}$ = 100 nm. (c) The electric field component $\sqrt{|\vec{E_x}|^2 + |\vec{E_z}|^2}$ distribution of the quasi-TE modes[47] for the optimized silicon-graphene hybrid plasmonic waveguide (@ 2 μm). (d) Calculated graphene absorptance $\eta$ as the propagation length L varies for the cases with different metal widths $w_m$ = 100, 200, and 300 nm. Here $h_m$ = 50 nm, $w_{si}$ = 3 μm, and $h_{si}$ = 100 nm.

Fig. 2(b) shows the dependence of the ratio $\eta_g$ and the absorption coefficients $(\alpha_g, \alpha_m)$ on the width $w_m$ and the height $h_m$ of the metal strip. Here the dimensions of the silicon ridge are $w_{si}$ = 3 μm, and $h_{si}$ = 100 nm. It can be seen that a high ratio $\eta_g$ can be achieved by choosing a narrow metal strip, which is simply owing to significant reduction of the metal absorption. For example, when choosing $w_m$=100 nm, the metal absorption coefficient is as small as $\alpha_m$= 0.019 dB/μm while the ratio $\eta_g$ is even as high as ~90%. However, the graphene absorption coefficient $\alpha_g$ also decreases in some degree when the metal strip becomes narrow. Therefore, in order to have sufficiently high graphene absorption coefficient as well as high absorption ratio $\eta_g$, we choose $w_m$=200 nm in our design, which also makes the fabrication relatively easy and guarantees a low graphene-metal contact resistance for the middle electrode. The absorption coefficients ($\alpha_g$, $\alpha_m$) can also be enhanced further when reducing the metal thickness, as shown in Fig. 2(b). However, the graphene absorption ratio $\eta_g$ also decreases. Therefore, we choose $h_m$=50 nm by making a trade-off.

For the designed silicon-graphene hybrid plasmonic waveguide with the following parameters: $w_m$=200 nm, $h_m$= 50 nm, $w_{si}$= 3μm, and $h_{si}$= 100 nm, the calculated electric field distribution $\sqrt{|\vec{E_x}|^2 + |\vec{E_z}|^2}$ of the quasi-TE mode is shown in Fig. 2(c). It can be seen that there is strong filed localization and enhancement in the area around the metal strip. For example, the electric field component $\sqrt{|\vec{E_x}|^2 + |\vec{E_z}|^2}$ along the graphene layer at the metal corners reaches up to $1.0\times10^7$ V/m for 1 mW optical power, which helps enhance the light absorption in graphene. For the present design, we calculate the total graphene absorption $\eta(L)$ as the propagation distance $L$ varies from 0 to 50 μm, as shown in Fig. 2(d). It can be seen that the total graphene absorption is almost saturated to be about 68.6% for the case of $w_m$= 200 nm when the length $L$ is 50 μm. When the metal width has some deviation, e.g., $w_m$= 300 nm, the total graphene absorption is close to a saturated value of 51.4% when the length $L$ is 20 μm, which is because the metal absorption increases. In contrast, when $w_m$= 100 nm, the total graphene absorption increases to 78.7% (not saturated yet) when the length $L$ increases to 50 μm, which is due to the relatively low absorption coefficients ($\alpha_g$, $\alpha_m$). With such a design, the present silicon-graphene hybrid plasmonic waveguide achieves the best result compared to those reported silicon-graphene hybrid waveguides (which were developed for 1.55 μm). In order to achieve a direct comparison, the silicon-graphene hybrid plasmonic waveguide was also designed optimally for 1.55 μm (see Supplementary Section 1), and the graphene absorptance at 1.55 μm is about 54.3% for the optimal design with $w_m$=200 nm when the length $L$=20 μm. In contrast, in ref. 26 the graphene absorptance is 44% only even for the *bilayer*-graphene hybrid plasmonic waveguide with $w_m$=180 nm and $L$=22 μm. For the $Si_3N_4$-graphene hybrid plasmonic waveguide with $w_m$=70 nm in ref. 29, the graphene absorptance $\eta$ is 42% when the length $L$=40 μm. More recently a plasmonic-enhanced graphene waveguide with bowtie-shaped metallic structures was reported with a short device length of 6 μm, and however the graphene absorption is saturated to be ~34%[21].

## 3. Results and analyses.

The designed waveguide photodetectors were fabricated with a series of steps (see Method), including the processes of electron-beam lithography, ICP etching, $Al_2O_3$ atom-layer deposition, graphene transfer, metal deposition, etc. For the fabricated devices, the I-V characteristics were characterized by varying the gate voltage (see Supplementary Section 2.1). The contact resistance and the graphene properties were obtained from the data fitting of the measured resistances with a simple capacity model[31]. The typical value for the graphene mobility is ~500 $cm^2$/V·s. For all the devices, the total contact resistances are typically several tens of Ohms, depending on the sizes of the contact regions and some random variations introduced in the fabrication processes. As an example, the total contact resistance is about 45 Ω for Device A, which is characterized in more details in the following parts.

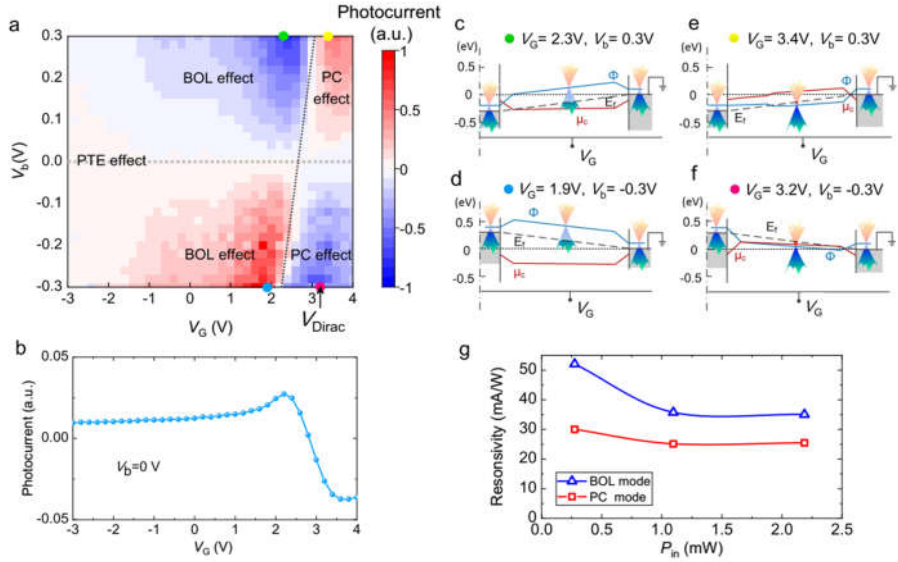

Fig. 3. Characterization and analysis for Device A. (a) Measured photocurrent map as the gate voltage $V_G$ and the bias voltage $V_b$ varies; (b) Dependence of the photocurrent at the zero bias on the gate voltage $V_G$; (c)-(f) Calculated Energy band diagrams for the cases of ($V_G$, $V_b$)=(2.3, 0.3), (1.9, −0.3), (3.4, 0.3), and (3.2, −0.3) V; (g) Measured responsivities with different input optical power $P_{in}$. Here $V_b$=−0.3V, and $V_G$=~1.9V (for the BOL effect), or $V_G$=~3.2V (for the PC effect).

The photocurrents were measured by using a lock-in amplifier (see Methods and Supplementary Section 6). The gate voltage $V_G$ is set to be less than 4.0 V in order to avoid the breakdown of the $Al_2O_3$ nano-layer. Fig. 3 (a) shows the measured photocurrent map for one of the representative devices (Device A) operating with different gate voltages $V_G$ and bias voltages $V_b$. For Device A, the Dirac voltage $V_{Dirac}$ is around 3.2 V (see the measurement in Supplementary Section 2.1). It can be seen that the photocurrent map has a 4-fold pattern, which is similar to the measured results for the device reported in ref. 30, even though the structural designs of the devices are different. From this figure, it can be seen that the photocurrent strongly depends on the gate voltages $V_G$ as well as the bias voltages $V_b$. In order to see more details, the dependence of the photocurrent at the zero bias on the gate voltage $V_G$ are shown in Fig. 3(b), which shows that there is a transition from a positive photocurrent to a negative one when the gate voltage $V_G$ is around 2.7 V. As it is well known, such a behavior for the dependence of the photocurrent on the gate voltage $V_G$ is very typical for the PTE photocurrent[48,49]. Our simulation with the photocurrent modeling in Supplementary Section 5 [see Fig. S7(d)] further confirms that the PTE effect is the dominant mechanism for the zero-bias photocurrent. As shown by the 4-fold pattern of Fig. 3(a), when the bias voltage $V_b$ is applied, the photocurrent increases greatly, which indicates the PTE effect is no longer the dominant mechanism. The reason is that the PTE photocurrent is generally not sensitive to the bias voltage $V_b$, as observed previously[31]. This is also predicted by the theoretical modeling in Supplementary Section 5. Instead, the dominant mechanisms for generating the photocurrent is very likely to be the BOL effect or the PC effect when $V_b \neq 0$. As shown in Fig. 3(a), the 4-fold photocurrent map has two sub-parts, i.e., the left and the right regions divided by the dotted line locating around $V_G$=2.3~3 V. At the left side, the measured photocurrent and the bias voltage have opposite signs, which indicates the dominant mechanism is the BOL effect[50]. In contrast, at the right side, the signs for the photocurrent and the bias voltage are consistent, which indicates the dominant mechanism is the PC effect[31].

In order to better understand the mechanisms of the photodetectors, we also give theoretical calculations for the Fermi level $E_f$, the Dirac-point energy $\Phi$, and the chemical potential $\mu_c$ along the graphene channel between the signal-electrode and the right ground-electrode (see the details in Supplementary Section 4), as shown in Fig. 3(c)-(f). In this calculation, the bias voltage is chosen to be $V_b = \pm 0.3$ V while the gate voltage is chosen as $V_G = \sim 2.0$ V and $\sim 3.2$ V to be locating at the left and right sides of the photocurrent map [see the labels in Fig. 3(a)]. Here the chemical potential for the graphene sheet underneath the gold electrodes is estimated to be around −0.1 eV due to the pinning effect[51]. In contrast, the chemical potential of the graphene sheet in the channel center is fully gate-controllable, and there is a transition region gradually varying from the pinning region and the fully gate-controllable region. As shown in Fig. 3(c)-(d), which are respectively for the cases with $(V_G, V_b)$ = (2.3, 0.3) V and (1.9, −0.3) V, the graphene sheet is highly doped. As a result, the bolometric coefficient $\beta$ is large[13, 50] and thus the BOL effect becomes the dominant mechanism. In Fig. 3 (e)-(f), which are respectively for the cases with $(V_G, V_b)$ = (3.4, 0.3) V and (3.2, −0.3) V, the graphene sheet is lightly doped. As a result, the bolometric coefficient $\beta$ becomes small[13,50] and thus the BOL effect is suppressed. Meanwhile, the lifetime of photo-generated carriers in graphene becomes long because of the low doping level[50]. In this case, the density of the photogenerated carrier is sufficiently high and the PC-effect becomes the dominant mechanism for photoresponse.

In summary, when the bias voltage $|V_b|$ increases from 0 to 0.3 V, the dominant mechanism for the photoresponse changes from the PTE effect to the BOL effect or the PC effect, depending on the applied gate voltage. Meanwhile, the responsivity increases significantly if the gate voltage is controlled well. Fig. 3(g) shows the measured responsivity for Device A operating with $V_b$ = −0.3 V when choosing $V_G$ = ~1.9 V (the BOL effect), and ~3.2 V (the PC effect), respectively. It can be seen that the responsivities for the BOL and PC modes are 35.0 mA/W, and 25.5 mA/W, respectively, when the input optical power $P_{in}$ is ~2.2 mW. When the input optical power $P_{in}$ decreases to 0.28 mW, the responsivities increase to about 52.1 mA/W and 30.0 mA/W for the BOL mode ($V_G$ = ~1.9 V) and the PC mode ($V_G$ = ~3.2 V), respectively.

The frequency responses of the devices were measured by using a setup combining a commercial 10 GHz optical modulator and a vector network analyzer (VNA, 40 GHz bandwidth), as shown in Fig. 4(a)-(b). The gate voltages were chosen as $V_G$ = 2.1 V and 3.4 V, corresponding to the BOL effect and the PC effect, respectively. Because the output optical power of the optical modulator at 2 μm is limited and there is no 2 μm optical amplifier available in the lab, the input optical power to the photodetectors is limited to 0.5 mW. In this case, the small-signal photocurrent (in the scale of µA) is much lower than the dark current (~3 mA) and thus some notable noise was observed at high frequencies in the measurement, as shown in Fig. 4(a)-(b). From this figure, no notable decay is observed in the frequency range of 1.5~20 GHz for both cases with the BOL effect and the PC effect. Here the maximal frequency in the measurement is up to 20 GHz, limited by the 2 µm optical modulator available in the lab. In order to estimate the 3 dB-bandwidth of the present photodetectors, an equivalent circuit is established (see Supplementary Section 2.2) and the parameters for all the RCL elements in the circuit are extracted by fitting the measured data of $S_{11}$. The established equivalent circuit was then used to estimate the 3 dB-bandwidth of Device A. As shown in Fig. 4, the estimated 3 dB-bandwidths $BW_{3dB}$ are about 110 GHz and 92 GHz for the BOL effect and the PC effect, respectively.

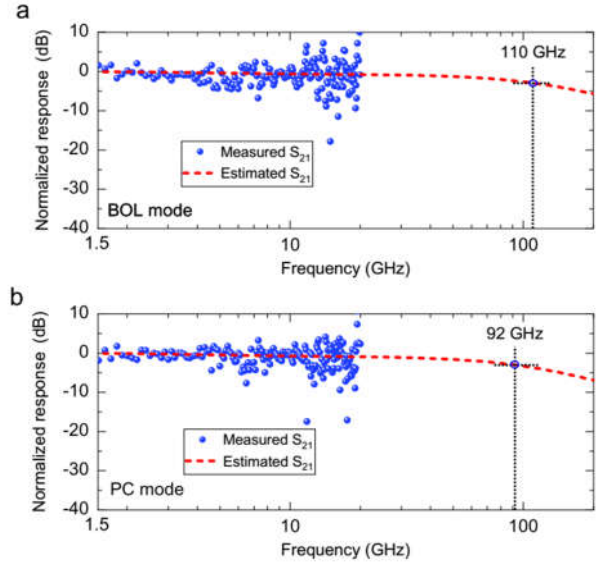

Fig. 4. Measured frequency responses and estimtaed $S_{21}$ from the equivalent circuits of Device A operating at different gate voltages when choosing $V_b = -0.3$ V: (a) the BOL mode ($V_G$=2.1 V); (b) the PC mode ($V_G$=3.4 V).

Fig. 5(a)-(b) show the measured responsivity and the frequency response for another photodetector (Device B) on the same chip. For Device B, the graphene is highly p-doped with a Dirac voltage $V_{Dirac}$ lager than 4.0 V [see Fig. S3 (a)], which is the maximal gate voltage used in our experiment regarding the breakdown condition of the 10 nm-thick $Al_2O_3$ layer. In this case, Device B works with the BOL effect. As shown in Fig. 5(a), the responsivity is up to 70 mA/W when $V_b$ =−0.3 V and $P_{in}$=0.28 mW. From the measured frequency response shown in Fig. 5(b),

the 3 dB-bandwidth $BW_{3dB}$ is over 20 GHz, while the 3 dB-bandwidth estimated from the equivalent circuit is as high as 123 GHz.

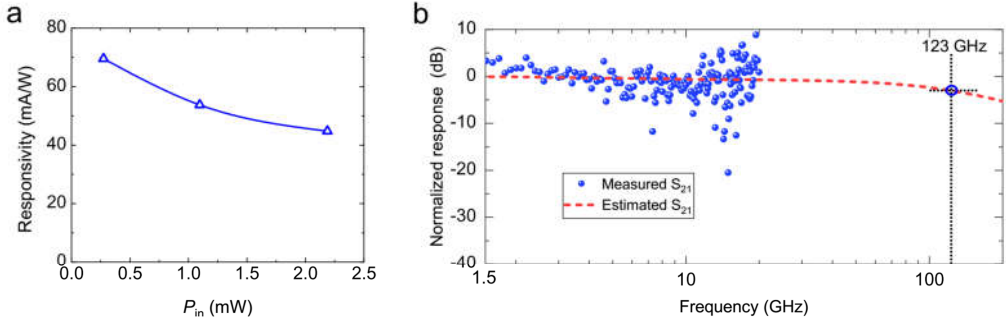

Fig. 5 Experimental results of Device B at 2 μm: (a) the measured responsivity as the input power $P_{in}$ varies ($V_b$= −0.3 V); (b) The measured frequency response ($V_b$= −0.5 V, $V_G$=2.9 V) and the estimated $S_{21}$ from the equivalent circuit.

In order to verify the high-bandwidth of the present waveguide photodetector, we characterized another device (Device C) on the same chip, as shown in Fig. 6(a). Device C is very similar to Devices A & B, while it has the grating coupler for 1.55 μm, so that the high-speed measurement setup for 1.55 μm available in the lab can be used. For Device C with a 20 μm-long absorption length, the Dirac voltage $V_{Diarc}$ is higher than 4 V [see Fig. S3 (a)], and the BOL effect is the dominant mechanism. From Fig. 6(a), Device C has a responsivity of 396 mA/W when $V_b$ = −0.3 V and $P_{in}$=0.16 mW. The high responsivity of Device C is attributed to higher light absorption in graphene and thus the higher light-induced temperature-increase which is beneficial to achieve high bolometric photoresponse. Fig. 6(b) shows the measured frequency response of Device C operating at $V_b$=0.6 V, which was characterized with the help of an Erbium-doped fiber amplifier @1.55 μm. It can be seen that the noise is low and the measured 3 dB-bandwidth is higher than 40 GHz (which is the maximal bandwidth of our VNA). Again, we also estimated the 3 dB-bandwidth for Device C according to the established equivalent circuit. It can be seen that Device C has an estimated 3 dB-bandwidth of 94 GHz. This device was further used to receive high bit-rate data with the setup shown in Fig. S8(d). Fig. 6(c) show the measured eye-diagram for the photodetector operating at 30 Gbit/s when $V_b$=0.6 V and $V_G$=2.8 V. It can be seen that the eye-diagram is open with a bit rate as high as 30 Gbit/s. More details are given in Supplementary

Section 6.

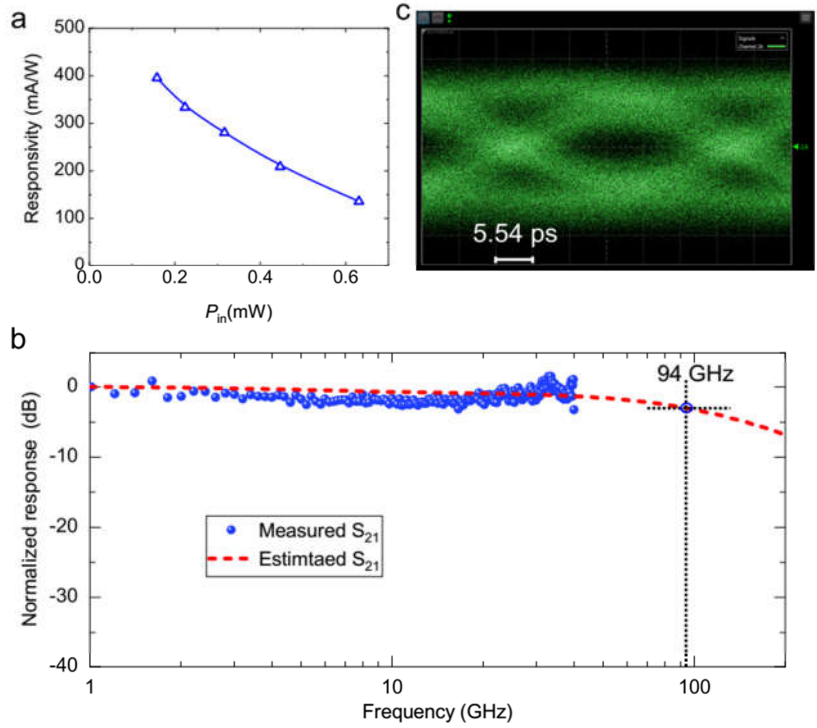

Fig. 6 Experimental results of Device C at 1.55 μm. (a) Measured responsivities at $V_b = -0.3$ V as the input optical power $P_{in}$ varies; (b) The measured frequency responses ($V_b = 0.6$ V, $V_G = 2.8$ V) and the estimated $S_{21}$ from the equivalent circuits; (c) Measured eye-diagram for a 30 Gbps PRBS data stream when operating with $V_b = -1$ V, $V_G = 0.3$ V.

4. **Discussions and outlooks**

Here we give a comprehensive comparion for the performces of the reported silicon-graphene photodetectors beyond 1.55 μm, as shown in Table 1. There have been several surface-illumiated silicon-graphene photodetectors with broad operation wavelength-bands. In ref. 37, a silicon-graphene photodetector was demonstrated with a responsivty of 6.25 mA/W @ 10 μm, and an estimated 3 dB-bandwidth of >1 GHz @ 1.03 μm. In ref. 38, a silicon-graphene photodetector was reported with responsivities of 0.6~0.076 A/W for an input optical power of 2.5~50 μW. For this device ref. 38, the measured 3 dB-bandwidth is higher than 50 GHz @ 0.8 μm, and the responsivity is 2~11.5 A/W for ultralow optical power in the wavelength range of 3~20 μm. For the waveguide photodetector reported recently[37, 43-45], the measured 3dB-bandwidths is in the scale of kHz or not given. In contrast, the present photodetectors (e.g., Device B) has a responsivity of 70 mA/W (@−0.3 V and 0.28 mW), and a set-up limited 3dB-bandwidth of > 20 GHz (the estimated 3 dB-bandwidth is ~124 GHz). To the best of our knowledgement, this is the first waveguide photodetectors with a 3dB-bandwidth of >20 GHz reported for the 2 μm wavelength-band.

Table 1. Performances of the graphene photodetectors for the mid-infrared range beyond the wavelength-band of 1.55 μm.

| Reference | Type | Mechanism | λ (μm) | (External) Responsivity | $P_{in}$ (μW) | $V_{bias}$ (V) | BW$_{3dB}$ |
|---|---|---|---|---|---|---|---|
| 37 | GSH, surface-illuminated | IPE | 2 | 0.16 mA/W | ~ 0.5 | 0 | ~ KHz |
| 38[a] | MGM, surface-illuminated | PV | 3 | 2 A/W | 2.5 | 0.02 | - |
| 39[b] | MGM, surface-illuminated (@T=10 K) | BOL | 10 | 6.25 mA/W | 0.8 | 2.4×10$^{-5}$ | - |
| 40 | graphene-barrier-graphene, surface-illuminated | Photogating | Up to 3.2 | >1 A/W | ~6 | 1 | ~Hz |
| 43 | GSH, waveguide-type | IPE | 2.75 | 130 mA/W | <1 | 1.5 | ~11 KHz |
| 44 | GSH, waveguide- type | IPE | 2.75 | 4.5 mA/W | 10 | -1 | - |
| 45 | MGM, waveguide- type | - | 3.8 | 2 mA/W | ~ 300 | -1 | - |
| This work: Device A | MGM, waveguide- type | BOL | 2 | 52 mA/W | 280 | -0.3 | >20 GHz[c] |
| This work: Device B | | | | 70 mA/W | | | >20 GHz[d] |

MGM: metal-graphene-metal (MGM).
GSH: graphene-semiconductor heterostructure.
Note a: The operation wavelength ranges from 0.8 μm to 20 μm, the 3 dB-bandwidth of 50 GHz was measured at λ=0.8 μm.
Note b: The operation wavelengths are 0.658, 1.03, 2, and 10 μm. External responsivity was evaluated from internal responsivity of 2×10$^5$ V/W, while the grapheme absorptance is 0.5%. The 3 dB-bandwidth of >1 GHz was measured at 1.03 μm.
Note c & d: The 3 dB-bandwidth are set-up limited. The estimated ones are 110 GHz and 123 GHz according to the equivalent circuits.

We further give a comparison for the reported silicon-graphene photodetectors at the wavelength-band of 1.55 μm because there are abundant measurement results reported, as shown in Fig. 7. Here only those devices with a monolayer graphene sheet and a 3dB-bandwidth of >1 GHz are included. It can be seen that a number of results for the realization of high bandwidths of > 40 GHz were reported[20-23, 28, 30-31, 33-35]. In particular, in ref. 22, the 3 dB- bandwidth is higher than 128 GHz. More recently, the device demonstraed in ref. 21 shows a 3 dB- bandwidth of over 110 GHz and the 100 Gbps data-receiving. Similarly, the present silicon- graphene hybrid waveguide photodetectors also demonstrate a high 3 dB-bandwidth of >40 GHz (which is setup-limited) and the estimated bandwidth from the equivalent circuit is around 100 GHz.

On the other hand, most of the reported graphene photodetectors have a responsivity of less than 100 mA/W[20, 23, 27-31, 33-35] when operating at a low bias voltage, e.g., $|V_b|$<0.3 V. As it is well known, for the metal-graphene-metal (MGM) photodetectors, the responsivity is usually in positive correlation to the bias-voltage $V_b$[20-23,25,27-32,37,38,45,50] and in negative correlation to the input optical power $P_{in}$[21,22,37,38]. Meanwhile, it is usually desired to be able to detect low optical power under a low bias voltage because a low bias voltage operation helps to reduce the dark currents and suppress the shot noise. In Fig. 7, the device responsivities are given under the bias voltages of $V_b=\pm 0.3$ V unless no data provided in the literatures. There are three graphene photodetectors with a responsivity of >100 mA/W reported recently[21,30,32]. For the one reported in ref. 30, the responsivitiy is estimated to ~150 mA/W (@ 0.3 V) with $P_{in}$=0.025 mW according to the given responsivities for the cases of $V_b$=0 and 1.2 V. The other one in ref. 32 has a responsivitiy of ~140 mA/W (@ 0.3 V) with $P_{in}$=0.56 mW, which is estimated from the given responsivities for the cases of $V_b$= 0 and 0.4 V. In ref. 21, the responsivities are proportional to the bias voltage and become ~375 mA/W and ~150 mA/W when operating with $V_b$= −0.3 V for $P_{in}$ of 0.08 mW and 0.6 mW, respectively. For the present photodetector (Device C) operating at a low bias voltage $V_b$=−0.3 V, the responsivity at 1.55 μm is as high as ~0.4 A/W @ $P_{in}$=0.16 mW, which is the highest one among various high-speed graphene photodetectors reported until now. In addition, the tunneling photodiode in ref. 36 with estimated bandwidth of 56 GHz was not included in Fig. 7 since it operates at large bias voltage of ~10V while the dark current can be kept in nA scale, therefore it can realize high on-off current ratio with responsivity of 240 mA/W @ $P_{in}$=0.42 mW. However, the large bias voltage is not CMOS-compatible and induces large power consumption and there is a trade-off between short carrier transit time and small deivce capacitance when choosing the tunneling layer thickness. As a summary, it can be seen that the present silicon-graphene waveguide photodetector works well with a high responsivity and a high bandwidth.

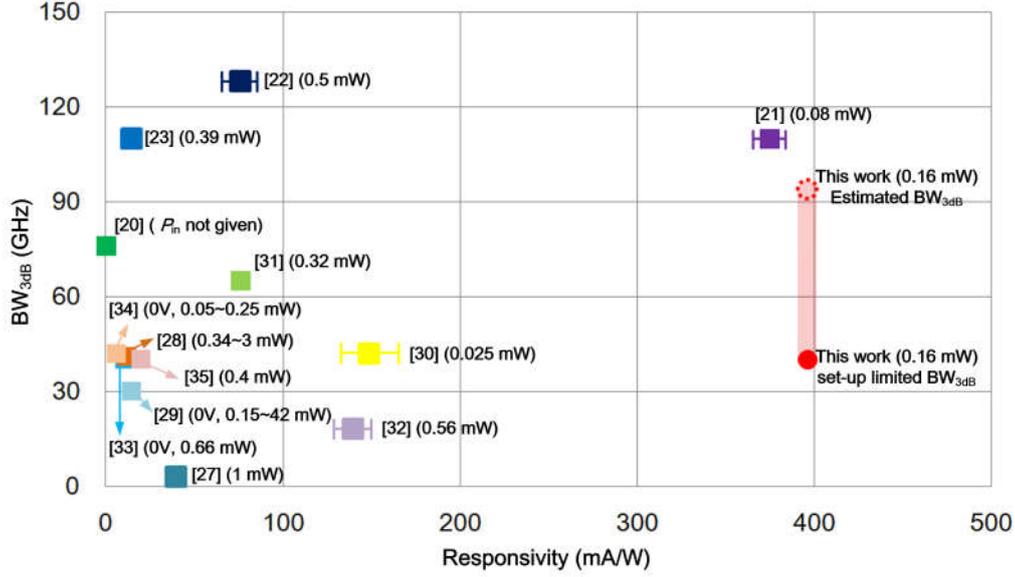

Fig. 7. Comprehensive comparisons for GHz graphene photodetectors at 1.55 μm reported previously. The data are with $V_b = \pm 0.3$V unless marked. In refs 21-23, 30, 32, the responsivites @ ± 0.3V are estimated according to the given data in the literatures.

5.  **Conclusions**

In this paper, we have proposed and demonstrated a novel silicon-graphene hybrid plasmonic waveguide photodetectors beyond 1.55 μm, which was realized by introducing an ultra-thin wide SOI ridge core region with a narrow metal cap at the top. With this design, the light absorption in graphene is enhanced while the metal absorption loss is reduced simultaneously. This greatly helps achieve effective optical absorption of graphene within a short length. The metal-graphene-metal sandwiched electrodes were also introduced to reduce the metal-graphene contact resistance, and thus helps improve the response speed. For the fabricated photodetectors, the mechanism has been revealed from the IV characteristics operating at different gate voltages. It has been shown that the dominant mechanism for the present photodetectors is the PTE effect at zero bias voltage, and becomes the BOL or PC effect at non-zero bias voltages, which helps achieve high-speed responses. For the fabricated photodetector operates at 2 μm, the measured 3 dB-bandwidth is >20 GHz (which is limited by the experimental setup), and the 3dB-bandwith estimated from the established equivalent circuit is as high as ~100 GHz, while the responsivity is ~70 mA/W for $P_{in}$=0.28 mW when operating at $V_b$=−0.3V. To the best of our knowledge, this is the first demonstration for 2 μm waveguide photodetectors with a 3dB-bandwidth of >20 GHz. In order to verify the ability for ultrafast photodetection, we have also measured the frequency responses for the present waveguide photodetector operating at 1.55 μm. It is shown that the measured 3 dB-bandwidth is >40 GHz (which is still limited by the setup) and the 3 dB-bandwith estimated from the equivalent circuit is as high as ~100 GHz. Meanwhile, the measured responsivity is about 0.4 A/W for an optical power of 0.16 mW when $V_b= -0.3$ V. As a summary, the present silicon-graphene photodetector is one of the best devices reported until now. The present work paves the way for achieving high-responsivity and high-speed near/mid-infrared waveguide photodetectors on silicon, which play an important role for various applications, including optical communications, nonlinear photonics, Lidar, on-chip spectroscopy, etc.

**Methods**

**Device Fabrication.** The ultrathin silicon core layer was obtained from a standard 220 nm-thick SOI wafer. A thermal oxidation process was used to obtain ~100 nm-thick silicon top-layer from standard 220 nm-thick lightly p-doped SOI wafer. Twice processes of EBL and ICP were used for the fabrication of the silicon ridge waveguide with a silicon thickness $h_{Si}=$ ~100 nm, an etching depth $h_{et}=$ ~50 nm, and a ridge width $w_{Si} = 3$ μm. A 90 nm-thick Aluminum gate-electrode (with the ohmic contact) was fabricated by utilizing the lift-off processes. A 10 nm-thick $Al_2O_3$ layer was deposited on the SOI ridge waveguide by using an atomic-layer deposition (ALD) process. The bottom-layer of the side ground-electrodes is made of 15/50 nm-thick Ti/Au hybrid thin films. Then a single-layer graphene sheet was transferred onto the chip, and patterned by the processes of the EBL and the ICP etching. Finally, a 50 nm-thick Au layer was deposited and patterned to form the narrow signal electrode and the top-layer of the side ground electrodes.

**Transfer process of graphene.** A 300 nm-thick film of PMMA was spin-coated on the CVD-graphene sheet grown on a copper thin film at 4000 rpm. The PMMA/graphene/copper film was floated on aqueous ammonium persulfate (60 mg/mL) to remove copper and rinsed in deionized water. Then it was transferred onto the chip. The graphene-covered chip was dried, baked, and soaked in acetone and rinsed with isopropanol.

**Device measurement.** The responsivities of the photodetectors were characterized by using the low frequency measurements. The continuous-wave light from the fiber laser was modulated with a frequency of 0.2 kHz by a chopper and then coupled to the optical waveguide by using an on-chip grating coupler. The photocurrent was then amplified and recorded by using a pre-amplifier and a lock-in amplifier [see Fig. S8(a)-(b) in Supplementary Materials]. The input optical power $P_{in}$ was estimated according to the measured coupling efficiency of the grating coupler (~ 10.5 dB @ 2 μm) and the power splitting ratio of the directional coupler (~1 dB @ 2 μm). More details about the optical power analysis are given in Supplementary Section 3.

**Acknowledgment**

We thank Dr. Liang Gao and Dr. Lei Wang for helpful suggestions for the device measurements. This project is supported by National Science Fund for Distinguished Young Scholars (61725503), National Natural Science Foundation of China (NSFC) (1171101320, 61775195), Zhejiang Provincial Natural Science Foundation (LZ18F050001, LD19F050001), and National Major Research and Development Program (No. 2018YFB2200200).

**Author contributions**

J. G. designed the device. J. L. performed the device fabrication with the assistance of C. L.. W. W. and Z. N. performed the $Al_2O_3$ deposition. J. L. and J. G. performed the dark current, low frequency, and high frequency measurements with the assistance of C. L. and Y. Y.. Z. F. and H. Y. contributed to the eye-diagram test. J. G. performed the simulations and modeling. J. G., J. L. and D. D. analyzed the simulation and experimental results. J. G., J. L. and D. D. wrote the manuscript. All the authors contributed to the discussions and the manuscript revisions. D. D. supervised the project.

**Additional information**

Supplementary information is available in the online version of the paper. Correspondence and requests for materials should be addressed to D. D.

**Competing financial interests**

The authors declare no competing financial interests.

**References**

1. Thomson, D. et al. Roadmap on silicon photonics. *J. Opt*. **18**, 073003 (2016).

2. Soref, R. Group IV photonics: enabling 2 μm communications. *Nat. Photon.* **9**, 358-359 (2015).

3. Lin, H. et al. Mid-infrared integrated photonics on silicon: a perspective. *Nanophotonics*, **7**, 393-420 (2017).

4. Sun, J. et al. Large-scale nanophotonic phased array. *Nature*. **493**, 195(2013).

5. Lavchiev, V. M. & Jakoby, B. Photonics in the mid-infrared: challenges in single-chip integration and absorption sensing. *IEEE J. Select. Topics Quantum Electron.* **23**, 452-463 (2017).

6. Hu, T. et al. Silicon photonic platforms for mid-infrared applications. *Photonics Res.* **5**, 417-430 (2017).

7. Chen, H. et al. 100-Gbps RZ data reception in 67-GHz Si-contacted germanium waveguide pin photodetectors. *J. Lightwave Technol.* **35**, 722-726 (2017).

8. Conley, B. R. et al. Temperature dependent spectral response and detectivity of GeSn photoconductors on silicon for short wave infrared detection. *Opt. Express.* **22**, 15639-15652 (2014).

9. Gassenq, A. et al. GeSn/Ge heterostructure short-wave infrared photodetectors on silicon, *Opt. Express.* **20**, 27297–27303 (2012).

10. Xu, S. et al. High-speed photo detection at two-micron-wavelength: technology enablement by GeSn/Ge multiple-quantum-well photodiode on 300 mm Si substrate, *Opt. Express.* **27**, 5798-5813 (2019).

11. Wang, R. et al. 2 μm wavelength range InP-based type-II quantum well photodiodes heterogeneously integrated on silicon photonic integrated circuits," *Opt. Express.* **23**, 26834-26841 (2015).

12. Ackert, J. J. et al. High-speed detection at two micrometres with monolithic silicon photodiodes. *Nat. Photon.* **9**, 393-396 (2015).

13. Xia, F. et al. Two-dimensional material nanophotonics. *Nat. Photon.* **8**, 899 (2014).

14. Koppens, F. H. et al. Photo-detectors based on graphene, other two-dimensional materials and hybrid systems. *Nat. Nanotechnol.* **9**, 780 (2014).

15. Romagnoli, M. et al. Graphene-based integrated photonics for next-generation datacom and telecom. *Nat. Rev. Mater.* **3**, 392 (2018).

16. Castellanos-Gomez, A. Black phosphorus: narrow gap, wide applications. *J. Phys. Chem. Lett.* **6**, 4280-4291 (2015).

17. Youngblood, N., Chen, C., Koester, S. J. & Li, M. Waveguide-integrated black phosphorus photodetector with high responsivity and low dark current. *Nat. Photon.* **9**, 247-252 (2015).

18. Huang, L. et al. Waveguide-Integrated Black Phosphorus Photodetector for Mid-Infrared
20

Applications. *ACS Nano* **13**, 913–921 (2019).

19. Yin, Y. et al. High-speed and high-responsivity hybrid silicon/black-phosphorus waveguide photodetectors at 2 μm, *Laser Photon. Rev.* 1900032, (2019).
20. Schall, D. et al. Graphene photodetectors with a bandwidth> 76 GHz fabricated in a 6 ″wafer process line. *J. Phys. D: Appl. Phys.* **50**, 124004 (2017).
21. Ma, P. et al. Plasmonically enhanced graphene photodetector featuring 100 Gbit/s data reception, high-responsivity and compact size. *ACS Photon.* **6**, 154-161 (2018).
22. Schall, D. et al. Record high bandwidth integrated graphene photodetectors for communication beyond 180 Gb/s, *Optical Fiber Communication Conf.* paper M2I. 4 (2018).
23. Ding, Y. et al. Ultra-compact graphene plasmonic photodetector with the bandwidth over 110GHz. *arXiv preprint arXiv:1808.04815* (2018).
24. Xia, F., Mueller, T., Lin, Y., Valdes-Garcia, A., & Avouris, P. Ultrafast graphene photodetector. *Nat. Nanotechnol.* **4**, 839–843 (2009).
25. Gan, X. et al. Chip-integrated ultrafast graphene photodetector with high responsivity. *Nat. Photon.* **7**, 883-887 (2013).
26. Pospischil, A. et al. CMOS-compatible graphene photodetector covering all optical communication bands. *Nat. Photon.* **7**, 892-896 (2013).
27. Youngblood, N. et al. Multifunctional graphene optical modulator and photodetector integrated on silicon waveguides. *Nano Lett.* **14**, 2741-2746 (2014).
28. Schall, D. et al.50 GBit/s photodetectors based on wafer-scale graphene for integrated silicon photonic communication systems. *ACS Photon.* **1**, 781-784 (2014).
29. Gao, Y. et al. High-performance chemical vapor deposited graphene-on-silicon nitride waveguide photodetectors. *Opt. Lett.* **43**, 1399-1402 (2018).
30. Shiue, R. et al. High-responsivity graphene–boron nitride photodetector and autocorrelator in a silicon photonic integrated circuit. *Nano Lett.* **15**, 7288 -7293 (2015).
31. Schuler, S. et al. Controlled generation of a p–n junction in a waveguide integrated graphene photodetector. *Nano Lett.* **16**, 7107-7112 (2016).
32. Schuler, S. et al. Graphene Photodetector Integrated on a Photonic Crystal Defect Waveguide. *ACS Photon.* **5**, 4758-4763 (2018).
33. Marconi, S. et al. Waveguide Integrated CVD Graphene Photo-Thermo-Electric Detector With> 40GHz Bandwidth. *Conf. Lasers and Electro-Optics* paper STh4N. 2 (2019).
34. Muench, J. E. et al. Waveguide-integrated, plasmonic enhanced graphene photodetectors. *arXiv preprint arXiv:1905.04639* (2019).
35. Li, T. et al. Spatially controlled electrostatic doping in graphene pin junction for hybrid silicon photodiode. *npj 2D Materials and Applications* **2**, 36 (2018).
36. Gao, Y. et al. High-speed van der Waals heterostructure tunneling photodiodes integrated on silicon nitride waveguides. *Optica*, **6**, 514-517(2019).
37. Casalino, M. et al. Free-Space Schottky Graphene/Silicon Photodetectors Operating at 2 μm. *ACS Photon.* **5**, 4577−4585 (2018).

38. Cakmakyapan, S., Lu, P. K., Navabi, A. & Jarrahi, M. Gold-patched graphene nano-stripes for high-responsivity and ultrafast photodetection from the visible to infrared regime. *Light-Sci. Appl.* **7**, 20 (2018).
39. Yan, J. et al. Dual-gated bilayer graphene hot-electron bolometer. *Nat. Nanotechnol.* **7**, 472–478 (2012).
40. Liu, C-H., Chang, Y., Norris, T. B. & Zhong, Z. Graphene photodetectors with ultra-broadband and high responsivity at room temperature. *Nat. Nanotechnol.* **9**, 273–278 (2014).
41. Badioli, M. et al. Phonon-Mediated Mid-Infrared Photoresponse of Graphene. *Nano Lett.* **14**, 6374-6381 (2014).
42. Freitag, Marcus. et al. Substrate-Sensitive Mid-infrared Photoresponse in Graphene. *ACS Nano* **8**, 8350-8356 (2014).
43. Wang, X. et al. High-responsivity graphene/silicon-hetero- structure waveguide photodetectors. *Nat. Photon.* **7**, 888-891 (2013).
44. Cheng, Z. et al. Graphene on silicon-on-sapphire waveguide photodetectors. *Conf. Lasers and Electro-Optics* paper STh1I. 5 (2015).
45. Qu, Z. et al. Waveguide integrated graphene mid-infrared photodetector. *Proc. SPIE Silicon Photonics XIII* **10537**, 105371N (2018).
46. Franklin, A. D. et al. Double contacts for improved performance of graphene transistors. *IEEE Electron Device Lett.* **33**, 17-19 (2012).
47. Guo, J., Wu, W. & Zhao, Y. Enhanced light absorption in waveguide Schottky photodetector integrated with ultrathin metal/silicide stripe, *Opt. Express.* **25**, 10057-10069 (2017).
48. Tielrooij, K. J. et al. Hot-carrier photocurrent effects at graphene-metal interfaces. *J. Phys.: Condens. Matter.* **27**, 164207 (2015).
49. Ma, Q. et al. Competing channels for hot-electron cooling in graphene. *Phys. Rev. Lett.* **112**, 247401 (2014).
50. Freitag, M., Low, T., Xia, F. & Avouris, P. Photoconductivity of biased graphene. *Nat. Photon.* **7**, 53 (2013).
51. Varykhalov, A., Scholz, M. R., Kim T. K. & Rader, O. Effect of noble-metal contacts on doping and band gap of graphene. *Phys. Rev. B*. **82**, 121101 (2010).

# Supplementary Materials

1 **Characteristic analysis of the silicon-graphene hybrid plasmonic waveguide.**

For the mode analysis of the silicon-graphene hybrid plasmonic waveguide, an FEM mode-solver from COMSOL was used. In this calculation, graphene is incorporated as a surface current boundary between two regions of dielectrics[1]. The optical conductivity of graphene was calculated from the Kubo formula[2], as shown in Fig. S1. In the simulations, we set the graphene chemical potential $\mu_c$ to −0.1 eV, in which situation $V_b$= 0, $V_G$= $V_{Dirac}$−0.15 V. Correspondingly the real parts of the optical conductivities basically are $\sigma_0$= 60.8 mS for both 2 μm and 1.55 μm (see Fig. S1).

When the voltages $V_G$ and $V_b$ vary, the chemical potential and the optical conductivity distributions for the graphene sheet change, while the mode field distributions change very slightly. It is noted that the light absorption of graphene is mainly decided by the real part (rather than the imaginary part) of its optical conductivity, and the light absorption mainly happens in the area close to the signal-electrode at the middle. Therefore, the real part of the optical conductivity of the graphene in the area close to the signal electrode is the key factor. Due to the pinning effect[3], the chemical potential $\mu_c$ of the graphene sheet underneath the signal-electrode (gold) is usually fixed to −0.1 eV according to the result in ref. 4. For the graphene sheet around the signal-electrode (gold), the chemical potential $\mu_c$ deviates slightly from −0.1 eV even for varied voltages $V_G$ and $V_b$. Meanwhile, the real part of the graphene optical conductivity varies very slightly for both wavelengths of 2 μm and 1.55 μm when $|\mu_c|$<0.2 eV, as shown in Fig. S1(a)-(b). Therefore, the simulation results in Fig. S1(a)-(b) for the case of $V_b$=0, $V_G$=$V_{Dirac}$−0.15 V are still valid even for varied voltages $V_b$ and $V_G$.

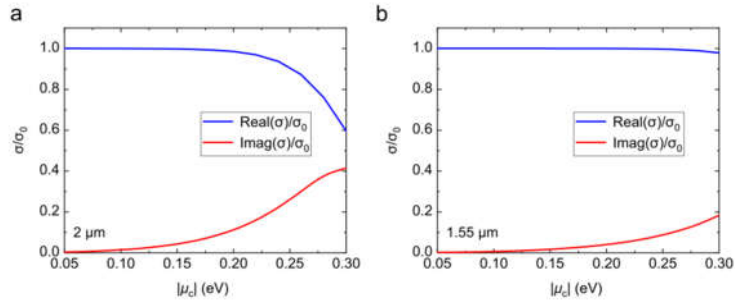

Fig. S1. The graphene conductivity versus the chemical potential for (a) 2 μm and (b) 1.55μm.

As described in the main text, for a given propagation length of $L$, the graphene light absorptance can be expressed by $\eta(L) = \eta_g(1 - e^{-\alpha_e L})$, in which $\eta_g$ is the graphene absorption ratio, and $\alpha_e$ is the mode absorption coefficients in μm$^{-1}$. One has $\alpha_e = \alpha/4.34$, where $\alpha$ is the waveguide loss in dB/μm. The graphene absorptance is calculated by[1]

$$\eta(L) = \frac{\int_0^L \int A_g(l) e^{-\alpha_e z} dl dz}{P_0} = \frac{1}{\alpha_e}(1 - e^{-\alpha_e L}) \frac{\int A_g(l) dl}{P_0}, \quad (S1)$$

where $P_0$ is the input mode power, $l$ is the coordinate of the line integral along the graphene surface in the $xy$ plane, $A_g(l)$ is the graphene light absorption intensity. One has $A_g(l) = \frac{1}{2}\text{Real}(\sigma_g)|\vec{E}_t(l)|^2$ (W/m$^2$), where Real($\sigma_g$) is the real part of the graphene conductivity, $\vec{E}_t$ is the transverse component of the electric fields along the graphene surface of the launched waveguide mode (at $z = 0$). Similarly, the metal absorptance is calculated by[5]

$$\eta_m(L) = \frac{\int_0^L \iint A_m(x,y) e^{-\alpha_e z} dx dy dz}{P_0} = \frac{1}{\alpha_e}(1 - e^{-\alpha_e L}) \frac{\iint A_m(x,y) dx dy}{P_0}. \quad (S2)$$

Here the integral area of the $xy$-plane surface integral is in the metal area, and $A_m(x, y)$ is the metal absorption intensity given by $A_m(x,y) = \frac{1}{2}\omega \cdot \text{Imag}(\varepsilon_m)|\vec{E}(x,y)|^2$ (W/m$^3$), where $\omega$ is the angular optical frequency, Imag($\varepsilon_m$) is the imaginary part of the metal permittivity, $\vec{E}$ is the electric fields in the metal area. One has $\alpha_e = \frac{\int A_g(l) dl}{P_0} + \frac{\iint A_m(x,y) dx dy}{P_0}$ according to Eqs. (S1) and (S2). The graphene absorption coefficient $\alpha_{eg}$ and the metal absorption coefficient $\alpha_{em}$ are given as (in μm$^{-1}$)

$$\alpha_{eg} = \frac{\int A_g(l) dl}{P_0}, \quad (S3)$$

$$\alpha_{em} = \frac{\iint A_m(x,y) dx dy}{P_0}. \quad (S4)$$

Then one has $\alpha_e = \alpha_{eg} + \alpha_{em}$, and the graphene absorption ratio is given by $\eta_g = \frac{\eta(L)}{\eta(L) + \eta_m(L)} = \frac{\alpha_{eg}}{\alpha_{eg} + \alpha_{em}}$. Since $\alpha_g = 4.34\alpha_{eg}$ and $\alpha_m = 4.34\alpha_{em}$, one has $\eta_g = \frac{\alpha_g}{\alpha_g + \alpha_m}$. With these formulas, the absorption coefficients ($\alpha_g$, $\alpha_m$) and the graphene absorption ratio $\eta_g$ can be calculated as the waveguide dimensions varies. The calculation results for the silicon-graphene hybrid waveguide operating at 2 μm are given in Fig. 2 in the maintext. We also give an analysis for the same waveguide operating at 1.55 μm, as shown in Fig. S2. It can be seen the waveguide designed for the wavelength-band of 2 μm also works well for the wavelength-band of 1.55 μm. Fig. S2(a) shows the dependence of the graphene absorption ratio $\eta_g$ and the absorption coefficients ($\alpha_g$, $\alpha_m$) on the width $w_m$ and the height $h_m$ of the metal strip when $w_{si}$ = 3 μm and $h_{si}$ = 100 nm. When the metal strip becomes wider, the graphene absorption coefficient $\alpha_g$ is higher and the graphene ratio $\eta_g$ becomes lower. On the other hand, the graphene absorption coefficient $\alpha_g$ is lower and the graphene ratio $\eta_g$ becomes higher when choosing a thicker metal strip.

The waveguide structure with ($w_m$, $h_m$)=(200 nm, 50 nm) designed for the 2 μm wavelength-band also works for the 1.55 μm wavelength-band. In this case, one has ($\alpha_g$, $\alpha_m$) = (0.295, 0.181) dB/μm, and $\eta_g$=62.0%. Fig. S2(b) shows the transversal electric field distribution of the designed waveguide operating at 1.55 μm. The electric field component $\sqrt{|\vec{E_x}|^2 + |\vec{E_z}|^2}$ along the graphene layer at the metal corners reach up to 1.22×10$^7$ V/m for 1 mW input power. As shown in Fig. S2(c), the graphene absorptance of this designed waveguide is 54.3% when choosing the device length as short as 20 μm. When the metal width has some deviation to be e.g. $w_m$= 300 nm, the graphene absorptance is close to the saturated value of 42.1% for the length $L$ as short as 10 μm, which is due to the high absorption of the metal strip.

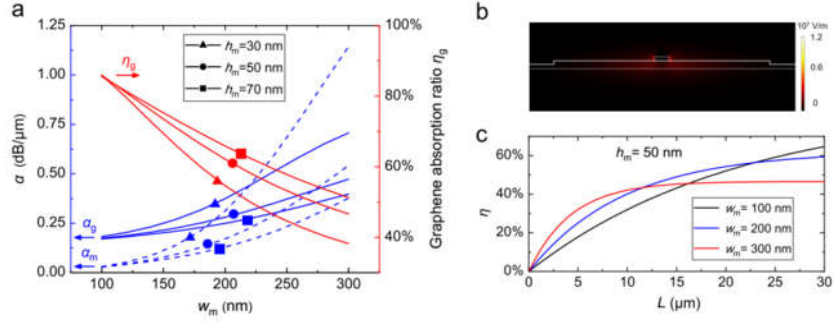

Fig.S2 (a) Calculated absorption coefficients ($\alpha_g$, $\alpha_m$), and the graphene absorption ratio $\eta_g$ as the metal-strip width $w_m$ varies for the cases with different metal heights $h_m$. Here $w_{si}$=3 μm, and $h_{si}$= 100 nm. (b) The electric field component $\sqrt{|\vec{E_x}|^2 + |\vec{E_z}|^2}$ of the quasi-TE mode for the optimized silicon-graphene hybrid plasmonic waveguide (@ 1.55 μm). (c) Calculated graphene absorptance $\eta$ as the propagation length $L$ varies for the cases with different metal-strip widths $w_m$ = 100, 200, and 300 nm. Here $h_m$= 50 nm, $w_{si}$= 3 μm, and $h_{si}$= 100 nm.

## 2   Device parameters.
2.1. Characterization of the silicon-graphene waveguide photodetectors.

For the fabricated devices, the I-V characteristics were characterized under varied gate voltages and a fixed low bias voltage supplied by two sourcemeters (Keithley 2401) in the dark case (i.e., the input optical power $P_{in}$=0). From the measured I-V curves, the total resistance R$_{tot}$ can be achieved easily. As it is well known, the total resistance for the photodetector based on a metal-graphene-metal structure is given by $R_{tot} = R_c + 0.5 \frac{W_g}{L_g}\sigma^{-1}$, where $R_c$ is the total contact resistance, $W_g$ and $L_g$ are respectively the width and the length of the graphene channel between the signal- and ground-electrodes, $\sigma$ is the graphene conductivity. The graphene conductivity is given as $\sigma = \sqrt{\sigma_{min}^2 + [\mu C_G(V_G - V_{Dirac})]^2}$, where $\sigma_{min}$ is the minimal conductivity, $\mu$ is the graphene mobility, $V_{Dirac}$ is the Dirac gate voltage, $C_G$ is the gate capacitance given by $C_G = \varepsilon_0\varepsilon_{Al_2O_3}/h_{Al_2O_3} = 8 \times 10^{-3} F/m^2$. Therefore, the contact resistance $R_c$ and the graphene properties ($\sigma_{min}$, $\mu$, and $V_{Dirac}$) can be obtained by fitting the measured data for the total resistances $R_{tot}$. For example, for Device A with $L_g$=50 μm and $W_g$=2.8 μm, the measured total resistance $R_{tot}$ is shown in Fig. 3(a). It can be seen that the total resistance becomes the maximal at $V_G$=3.2 V, which corresponds to the Dirac voltage. The other fitted parameters are $R_c$=45 Ω, $\sigma_{min}$=0.206 mS, and $\mu$=522 cm$^2$/V·s. For Device A, the minimum of the total resistance $R_{tot}$ is ~60 Ω when choosing $V_G$=−2 V, while the normalized contact resistance $R_c·L_g$ is about 2250 Ω·μm.

The measured data for another two devices (Devices B and C) are also given, as shown in Fig. S3(b). For these two devices, the graphene is highly doped, and their Dirac voltages $V_{Dirac}$ are larger than 4 V, and thus a low gate voltage (e.g., < 4V) does not introduce significant influence on the device resistances. Similarly, the contact resistance $R_c$ and the graphene properties ($\sigma_{min}$, $\mu$, and $V_{Dirac}$) for Devices B and C can also be obtained by fitting the measured data for the total resistances $R_{tot}$. For Device B with $L_g$=50 μm and $W_g$=2.8 μm and Device C with $L_g$=20 μm and $W_g$=2.2 μm (operating at 1.55 μm), their contact resistances $R_c$ are estimated as ~60 Ω and ~104 Ω, respectively. It can be seen that Devices A, B, and C have normalized contact resistances in the range of 2000~3000 Ω·μm. In our devices, the contact resistance $R_c$ depends on both the contact-resistances for the signal-electrode at the middle as well as the ground-electrodes at the sides. One should notice that the contact-resistances for the signal-electrode might be the dominant one because the signal-electrode is much narrower width than the ground-electrodes.

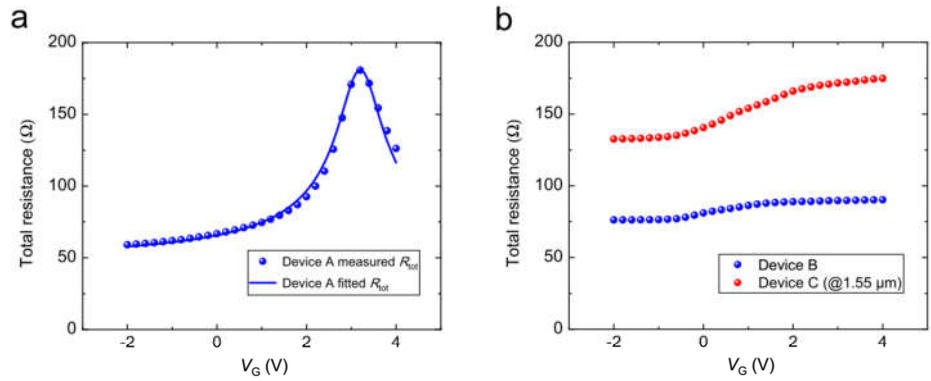

Fig. S3 (a) Measured total resistance $R_{tot}$ of Device A. (b) Measured total resistance $R_{tot}$ for Devices B and C.

2.2. Equivalent circuit for the photodetectors.

Since the 3 dB-bandwidth of the present photodetectors is beyond the setup-limit, here we establish an equivalent circuit model with the parameters extracted from the measured S$_{11}$, so that the frequency response S$_{21}$ can be calculated to estimate the 3 dB-bandwidth. Fig. S4 (a) shows the equivalent circuit, where $C_{pad}$ denotes the pad capacitance, $C_g$ and $R_g$ are respectively the capacitance and the resistance corresponding to the grapheme area, $C_{oxc}$ and $C_{oxs}$ are respectively the capacitances corresponding to the Al$_2$O$_3$ layer at the middle and the Al$_2$O$_3$ layer at the sides, $R_{Si}$ is the silicon resistance. According the structural symmetry of the device, this equivalent circuit model can further be simplified as shown in Fig. 4(b), where $C_{tot}=2(C_g+C_{pad})$, $C_{ox} = \frac{2C_{oxc}C_{oxs}}{C_{oxs}+C_{oxc}}$ and $R_{tot}= 0.5\ R_g$.

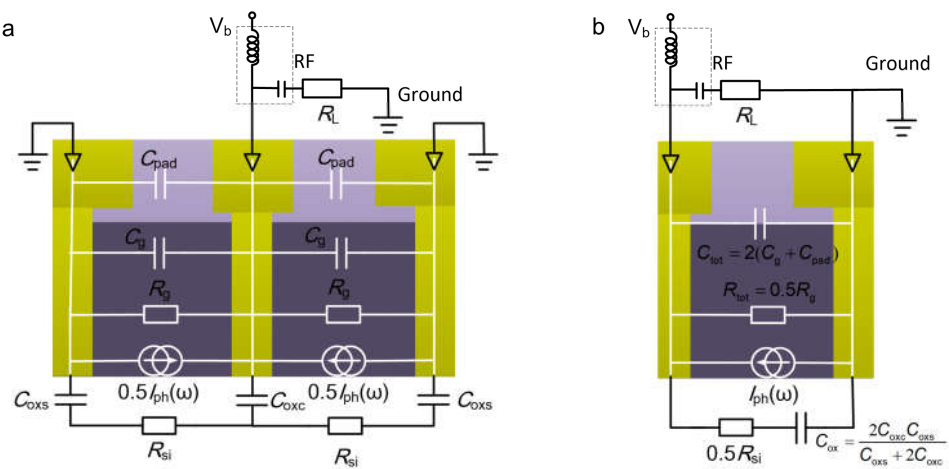

Fig. S4. (a) Equivalent circuit; (b) Simplified equivalent circuit model.

All these parameters can be extracted from the measured frequency-dependent impedance $Z_{in}$ given by $Z_{in}(\omega)=Z_0[1+S_{11}(\omega)]/[1-S_{11}(\omega)]$, where $Z_0=50\ \Omega$, the reflection coefficient $S_{11}$ was obtained by using a vector network analyzer (VNA). In particular, $S_{11}$ and $S_{21}$ were measured under the same gate voltages. Fig. S5(a)-(d) show the measured data for the impedances of the photodetectors when operating at different conditions. The parameters for the RC elements in the equivalent circuit were extracted by fitting the measurement data, as shown in Table S1. The fitted result for the total resistance given in Table S1 is very similar to the static measurement result (e.g., for Device A, $R_{tot}$=~98 Ω @$V_G$=2.1 V, ~150 Ω @ $V_G$=3.5 V), which indicates the established equivalent circuit works well for the present photodetectors. The impedances calculated from the equivalent circuit with the fitted parameters are also shown in Fig. S5(a)-(d), which further confirms that the established equivalent circuit model works well.

According the established equivalent circuit model, the frequency responses $S_{21}$ were calculated (see the results in Fig. 4 and Fig. 5 in the maintext) and the estimated 3 dB-bandwidths are listed in Table S1. For Device A, the estimated 3 dB-bandwidths are about 110 GHz and 92 GHz when operating with the BOL effect and the PC effect, respectively. The difference is mainly due to their different total resistances $R_{tot}$. As shown in Table S1, the total resistance $R_{tot}$ (98.6 Ω) under the BOL mode is lower than that (148.5 Ω) under the PC mode, which is due to the higher doping level in graphene. Accordingly, the estimated 3dB-bandwidth for Devices B is about 123 GHz, which is higher than that for Device C. According to the equivalent circuit model, the 3 dB-bandwidths of the present photodetectors are mainly determined by the total resistance $R_{tot}$ as well as the total capacitance $C_{tot}$. All the parameters are analyzed as below.

(1) About the total resistance $R_{tot}$. For Devices B and C, the lengths $L_g$ of their graphene sheets are respectively 50 μm and 20 μm, while the electrode spacings for them are 2.8 μm and 2.2 μm. As a result, Device C has a higher total resistance $R_{tot}$ than Device B.

(2) About the total capacitance $C_{tot}$. The total capacitance $C_{tot}$ consists of two parts. One is the capacitance $2 \cdot C_g$ for the graphene areas and the other one is the $2 \cdot C_{pad}$ for the metal pads/connectors. Here two-dimensional numerical simulations (using COMSOL Electrostatics Interface) were performed to evaluate the capacitances $2 \cdot C_g$ and $2 \cdot C_{pad}$. The simulation shows that the calculated normalized capacitances for the graphene areas of Devices B and C are about 32.1 pF/m and 34.5 pF/m. Accordingly, one has $2 \cdot C_g$=1.6 fF and 0.69 fF for Devices B and C regarding the length $L_g$=50 μm and 20 μm, respectively. The capacitance $2 \cdot C_{pad}$ consists of two parts contributed by the rectangular metal-pads and the metal-connectors. The capacitances contributed by the metal-connectors for all the devices are similar since their metal-connectors have similar shapes and sizes. For the present devices, the calculated results is ~6 fF capacitances. For the part contributed by the rectangular metal-pads, the calculated normalized capacitance is about 165 pF/m, which gives a capacitance of 26.4 fF for the 160 μm-long rectangular metal-pad accordingly. As a result, the capacitance $2 \cdot C_{pad}$ contributed by the rectangular metal-pads and the metal-connectors is about 32 fF for all the devices, which is much higher than the graphene capacitance $2 \cdot C_g$. Accordingly, all the devices with different lengths $L_g$ have similar capacitances $C_{tot}$ (~32 fF), which is consist with the fitted parameters (~40 fF) given in Table S1.

(3) About the silicon resistance $R_{si}$. The resistance $R_{si}$ depends on the device dimensions and the electrode layout. Even though Device C is shorter than Devices A and B, the resistances $R_{si}$ for these three devices are similar. One of the reasons is that Device C has narrower electrode-spacing than Devices A and B. Furthermore, the area of the signal-electrode sitting on the silicon layer for Device C is larger than that for Devices A and B, as shown in Fig. S5.

(4) About the $Al_2O_3$-layer capacitance $C_{ox}$. According to the device structure, the $Al_2O_3$-layer capacitance $C_{ox}$ is proportional to the overlap area between the electrode and the silicon core-layer. In Fig. S5(e)-(f), the detailed structures of Devices A & B and Device C are given by optical microscope pictures. For the present device, there are two parts contributed to the capacitance $C_{ox}$. One is for the signal-electrode at the middle and the other one is for the ground-electrodes at both sides. For Devices A and B [see Fig. S5(e)], the overlap area of the signal-electrode is much smaller than that of the ground-electrodes, and thus the capacitance $C_{oxc}$ corresponding to the signal-electrode is much smaller than the capacitance $C_{oxs}$ corresponding to the ground-electrodes, i.e., $C_{oxc} \ll C_{oxs}$. As a result, the capacitance $C_{oxc}$ is the dominant one. In contrast, for Device C, the area of the central electrode is actually similar to the ground-electrodes because there is a relatively large metal pad at the rear [see Fig. S5(f)]. On the other hand, the length of the ground-electrodes for Device C is shorter than that for Devices A and B. As a result, these devices have similar capacitances $C_{ox}$ while Device C is the smallest one, as shown in Table S1.

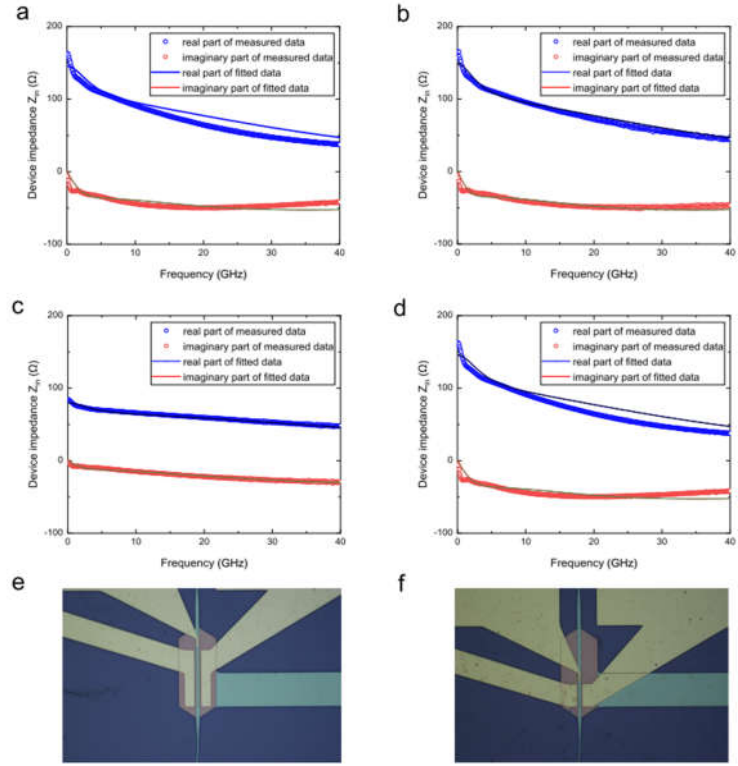

Fig. S5. The measured and fitted results for the impedances $Z_{in}$: (a) Device A at the BOL mode, $V_G$=2.1V. (b) Device A at the PC mode, $V_G$=3.4V. (c) Device B at the BOL mode, $V_G$=2.9V. (d) Device C at the BOL mode, $V_G$=2.8V. (e) Structure of Device A & B; (f) Structure of Device C.

Table S1. The parameters for the equivalent circuit extracted from the measured $S_{11}$ and the estimated 3dB-bandwdith $BW_{3dB}$ obtained from the frequency response calculated with the equivalent circuit.

| Devices | Mechanism | $R_{tot}$ (Ω) | $0.5 \cdot R_{si}$ (Ω) | $C_{tot}$ (fF) | $C_{ox}$ (fF) | $BW_{3dB}$ (GHz) | $L_g$ (μm) | $W_g$ (μm) | $\lambda$ (μm) |
|---|---|---|---|---|---|---|---|---|---|
| A | BOL effect | 98.6 | 370.8 | 39.7 | 95.2 | 110 | 50 | 2.8 | 2 |
|   | PC effect | 148.5 | 362.2 | 41.4 | 97.8 | 92 |  |  |  |
| B | BOL effect | 79.8 | 375.8 | 38.8 | 97.1 | 123 |  |  |  |
| C | BOL effect | 145.6 | 370.1 | 40.5 | 79.2 | 94 | 20 | 2.2 | 1.55 |

## 3  Characterization of the optical transmissions.

As shown in Fig. S6(a), light was coupled from an optical fiber to the waveguide photodetector by using a grating coupler. Here a directional coupler (DC) with a power splitting ratio of 90%:10% was inserted before the photodetector to make the fiber-alignment convenient. When operating at 2 μm, the fabricated grating coupler has a coupling loss of ~10.5 dB and the fabricated DC has a loss of ~1 dB. In contrast, when operating at 1.55 μm, the fabricated grating coupler has a coupling loss of about 8.5 dB and the fabricated DC has a loss of 1~2 dB (due to some unexpected fabrication variations).

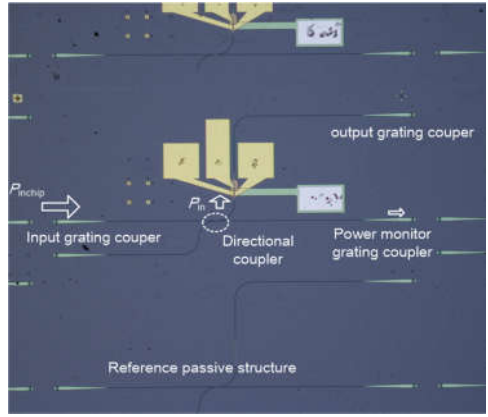

Fig. S6. Microscopy picture of the fabricated devices.

## 4  Calculations of the Fermi level, the chemical potential, and the Dirac-point energy.

The relationship for the Fermi level $E_F(x)$, the chemical potential $\mu_c(x)$, and the Dirac-point energy $\Phi(x)$ is described as

$$E_F(x) = \Phi(x) + \mu_c(x). \tag{S5}$$

Usually $E_F(x)$ is assumed to be varying linearly along the graphene channel[6], i.e.,

$$E_F(x) = -V_b(x) = V_b\left(\frac{x}{L_c} - 1\right). \tag{S6}$$

In this work, the width of the transition region between the pinning region and the fully gate-controllable region is around 0.3 μm[7]. In the graphene region unaffected by the metal pinning effect (i.e., 0.3 μm ≤ x ≤ $L_c$−0.3 μm), the electrostatic-doping induced charge-density $n(x)$ is given by

$$n(x) = \frac{C_G[V_G - V_b(x) - V_{Dirac}]}{e}, \tag{S7}$$

where $C_G$ is the capacitance. The charge density $n(x)$ is related with the chemical potential $\mu_c(x)$ by

$$n = n_e - n_h = \frac{2}{\pi(\hbar v_F)^2} \int_0^\infty \left(\frac{E}{e^{\frac{E-\mu_c}{k_BT}}+1} - \frac{E}{e^{\frac{E+\mu_c}{k_BT}}+1}\right) dE, \tag{S8}$$

where $n_e$ is the electron density, $n_h$ is the hole density, and $v_F$ is the graphene Fermi velocity (e.g., $10^6$ m/s here). Then the chemical potential $\mu_c(x)$ can be calculated from Eqs. (S7) and (S8). The chemical potential of the graphene underneath the gold electrode was set as −0.1eV (lightly p-doping)[4], and the chemical potential $\mu_c(x)$ in the pinning area ($0 \leq x \leq 0.3$ μm, $L_c - 0.3$ μm $\leq x \leq L_c$) is assumed to be varying linearly with the position $x$. Finally, the Dirac-point energy $\Phi(x)$ is calculated by Eq. (S5).

## 5  The PTE photocurrent modeling.

The PTE photocurrent $I_{\text{PTE}}$ is evaluated by $I_{\text{PTE}} = \frac{\bar{V}_{\text{PTE}}}{R_{\text{tot}}}$, where $\bar{V}_{\text{PTE}}$ is the average PTE photovoltage, $R_{\text{tot}}$ is the total resistance. The total resistance $R_{\text{tot}}$ was taken from the measured IV curves (see Supplementary Section 2.1). The average PTE photovoltage $\bar{V}_{\text{PTE}}$ was calculated with the formula $\bar{V}_{\text{PTE}} = -\int S(l) \frac{d\bar{T}_e(l)}{dl} dl$, where $l$ is the lateral position along the graphene sheet, $S(l)$ is the Seebeck coefficient and $\frac{d\bar{T}_e(l)}{dl}$ is the optically-induced electron temperature gradient. For the lateral positions at the corners of the signal electrode and the right ground electrode, one has $l=0$ and 2.8 μm for Device A, respectively. The Seebeck coefficient S is expressed as $S(\mu_c) = -\frac{\pi^2 k_B^2 T}{3e} \frac{1}{\sigma} \frac{d\sigma}{d\mu_c}$. The average electron temperature is given by $\bar{T}_e(l) = \frac{\int_0^L T_e(l,z) dz}{L}$, where $T_e(l, z)$ is the electron temperature distribution obtained from the following heat equation

$$-\nabla \cdot (\kappa \nabla T_e) + \frac{\kappa}{\zeta^2}(T_e - T_0) = P(l, z). \quad (S9)$$

In Eq. (S9), $T_0$=300 K, the cooling length $\zeta$ is chosen as $\zeta$=1 μm in graphene[8], and the electronic thermal conductivity $\kappa$ is given by Wiedeman-Franz relation $\kappa = \frac{\pi^2 k_B^2 T \sigma}{3e^2} = 2.44 \times 10^{-8} T\sigma \ [W/K]$. The optical power absorption density in Eq. (S9) is given by $P(l,z) = P_{\text{in}} \frac{A_g(l)}{P_0} 10^{-0.1\alpha z}$, where $P_{\text{in}}$ is the input power, and $\frac{A_g(l)}{P_0}$ is the normalized graphene absorption density in unit of m$^{-2}$ (see Supplementary Section 1) as shown in Fig. S7(a). One can see that most light absorption occurs near the interface between the graphene sheet and the signal electrode. It is noted that the light absorption distributions are not sensitive to $V_G$ and $V_b$ (−0.3~0.3V). Heat dissipation happens via the metallic contacts due to the heat-sink effect for the graphene underneath the metal electrodes[9], in which case the cooling length is reduced compared to the case for the pure graphene (without metals). Here we choose the cooling length as $\zeta_m$= 0.1 μm as an example, corresponding the case with strong heat-sink effect. In this case, the coefficient $\kappa/\zeta^2$ is 100 times higher than that for the case of ignoring the heat-sink effect and correspondingly the total power dissipation becomes 100 times higher. The calculated average electron temperature increment ($\bar{T}_e - T_0$) when $V_G$=$V_{\text{Dirac}}$ is shown in Fig. S7(a). One can see that the metal heat-sink effect introduces some degradation of the electron heating, while $[\bar{T}_e(l=0) - T_0]$ does not decrease to zero. The maximum value is achieved at the position close to the signal-electrode, i.e., the optically-induced electron temperature gradient $\frac{d\bar{T}_e(l)}{dl}$ is negative for most part of the graphene sheet. Fig. S7(b) shows the energy-band diagrams at the zero bias for the cases with different gate voltages $V_G$. For the graphene sheet, there are two transition regions and a fully gate-controllable region. When $V_b=0$, the working mechanism is possibly dominated by the PTE effect or the PV effect. If the mechanism is dominated by the PV effect, the photocurrent mainly depends on the build-in electric field $d\Phi/dl$ in the transition region close to the central metal strip. Since the build-in electric field changes monotonically as the gate voltage $V_G$ increases, as shown in Fig. S7(b), the photocurrent should change monotonically as the gate voltage $V_G$ increases, which however does not agree with the experimental result shown in Fig. 3(b). As a conclusion, the mechanism is not dominated by the PV effect.

Instead, the dominant mechanism is likely to be the PTE effect. Fig. S7 (c) shows the calculated PTE photo-current $I_{PTE}$ as the bias voltage $V_b$ and the gate voltage $V_G$ varies. From this figure, one sees that the photocurrent $I_{PTE}$ is highly dependent on the gate voltage. The zero point for the photocurrent $I_{PTE}(V_G)$ shifts as the bias voltage $V_b$ varies. The photocurrent $I_{PTE}$ is not sensitive to the bias voltage for fixed $V_G$ unless $|V_G-V_{Dirac}|<0.5V$. Fig. S7(d) shows the calculated photocurrent $I_{PTE}$ at zero bias as $V_G$ varies, which matches very well with the experimental result shown in Fig. 3(b). Therefore, it is verified that the dominant mechanism for the present photodetectors is the PTE effect when operating at zero bias.

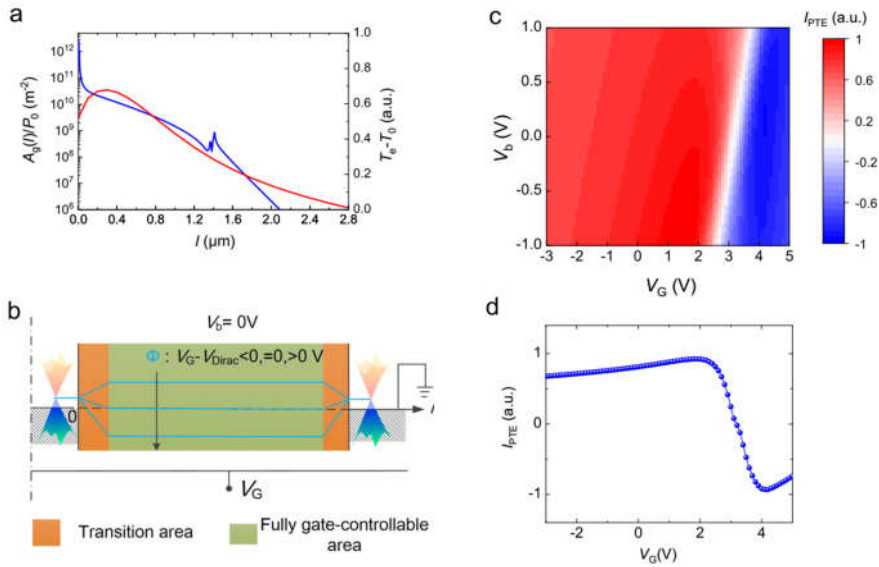

Fig. S7 Simulated results of Device A. (a) Normalized graphene absorption density $P(l)$ and average electron temperature increment ($\bar{T}_e - T_0$) with $V_G=V_{Dirac}$; (b) Energy-band diagram of the silicon-graphene photodetector with $V_b=0$; (c) PTE photo-current as the voltages $V_b$ and $V_G$ varies; (d) PTE photo-current versus $V_G$ when $V_b=0$.

## 6 Measurement setups

Fig. S8(a) shows the low-frequency measurement setups for the devices when operating at 2 μm. Here the CW light from the 2 μm fiber laser was modulated by a 0.2 kHz chopper. The modulated light was coupled to the chip through the input grating coupler for TE polarization. The polarization controller was used before light enters the chip. The bias voltage and the gate voltage were applied by Sourcemeters (Keithley 2401) and a pre-amplifier (SR570), respectively. The electrical signal was received by using a lock-in amplifier (SR830) with help of the reference clock signal from the chopper. The photocurrent from the lock-in amplifier was used to evaluate the responsivity. For the devices operating at 1.55 μm, light from the tunable laser (HP 8163A) was internally modulated with a frequency of 1 kHz, as shown in Fig. S8(b).

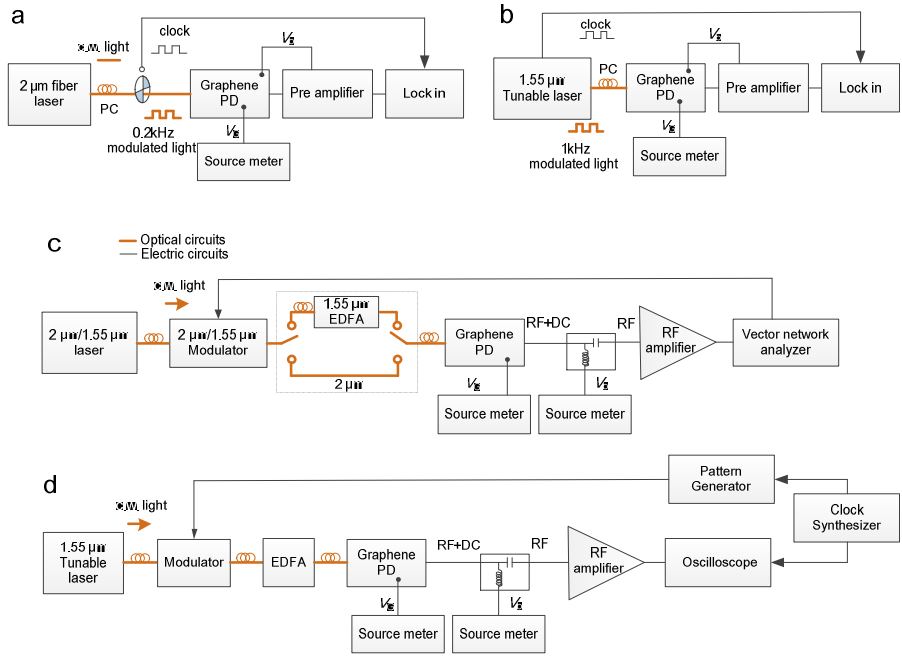

Fig. S8 (a) Low-frequency measurement setups for the devices when operating at 2 μm; (b) Low-frequency measurement setups for the devices when operating at 1.55 μm; (c) High-frequency measurement setups for the devices; (d) Eye-diagram measurement setups for the devices when operating at 1.55 μm.

Fig. S8(c) shows the experimental setup for measuring the high frequency response. The CW light was modulated with an optical modulator (2μm: IXBLUE MX2000-LN-10, 10 GHz bandwidth; 1.55 μm: Sumitomo T. MXH1.5DP-40PD-ADC, 22 GHz bandwidth), which was driven by the electrical signal from a vector network analyzer (ROHDE&SCHWARZ ZVA40, 40GHz). For the measurement of the 1.55 μm device, the modulated light was amplified by using an EDFA (Thorlabs EDFA100P) before it was coupled to the chip through an input grating coupler. For the measurement of the 2 μm device, no optical amplifier was available. The output electrical signal of the silicon-graphene waveguide photodetector was then amplified by using a RF amplifier (Centellax OA4MVM2) and finally received by the VNA. The calibrations were performed before the device tests with high-speed commercial photodetectors.

The eye diagram was measured with another test system, as shown in Fig. S8(d). The PRBS signal generated by a pattern generator (Keysight N4952A) was used for driving the optical modulator. The electrical signal output from the silicon-graphene photodetector was amplified by a microwave system amplifier (Keysight N4985A) and finally received by a wide-bandwidth oscilloscope (Keysight DCA-X 86100D). The clock signal generated by Keysight N4960A was used for the synchronization of the pattern generator and the oscilloscope.

6. **References**

1. Koester, S. J. & Li, M. Waveguide-coupled graphene optoelectronics. *IEEE J. Select. Topics Quantum Electron.* **20**, 84-94 (2014).
2. Hanson, G. W. Dyadic Green's functions and guided surface waves for a surface conductivity. *J. Appl. Phys.* **103**, 064302 (2008).
3. Huard, B. et al. Evidence of the role of contacts on the observed electron-hole asymmetry in graphene. *Phys. Rev. B*. **78**, 121402 (2008).
4. Varykhalov, A., Scholz, M. R., Kim T. K. & Rader, O. Effect of noble-metal contacts on doping and band gap of graphene. *Phys. Rev. B*. **82**, 121101 (2010).
5. Senanayake, P. et al. Surface plasmon-enhanced nanopillar photodetectors. *Nano lett.* **11**, 5279-5283 (2011).
6. Freitag, M., Low, T., Xia, F. & Avouris, P. Photoconductivity of biased graphene. *Nat. Photon.* **7**, 53 (2013).
7. Mueller, T. et al. Role of contacts in graphene transistors: A scanning photocurrent study. *Phys. Rev. B*. **79**, 245430 (2009).
8. Schuler, S. et al. Controlled generation of a p–n junction in a waveguide integrated graphene photodetector. *Nano Lett.* **16**, 7107-7112 (2016).
9. Low, T. et al. Cooling of photoexcited carriers in graphene by internal and substrate phonons. *Phys. Rev. B*. **86**, 045413 (2012).